# Dynamical charge density fluctuations pervading the phase diagram of a Cu-based high-$T_\text{c}$ superconductor

**Authors:** R. Arpaia[1,2], S. Caprara[3,4], R. Fumagalli[1], G. De Vecchi[1], Y.Y. Peng[1,†], E. Andersson[2], D. Betto[5], G. M. De Luca[6,7], N. B. Brookes[5], F. Lombardi[2], M. Salluzzo[7], L. Braicovich[1,5], C. Di Castro[3,4], M. Grilli[3,4], G. Ghiringhelli[1,8,*]

**Affiliations:**

[1] Dipartimento di Fisica, Politecnico di Milano, Piazza Leonardo da Vinci 32, I-20133 Milano, Italy

[2] Quantum Device Physics Laboratory, Department of Microtechnology and Nanoscience, Chalmers University of Technology, SE-41296 Göteborg, Sweden

[3] Dipartimento di Fisica, Università di Roma "La Sapienza", P.le Aldo Moro 5, I-00185 Roma, Italy

[4] CNR-ISC, via dei Taurini 19, I-00185 Roma, Italy

[5] ESRF, The European Synchrotron, 71 Avenue des Martyrs, F-38043 Grenoble, France.

[6] Dipartimento di Fisica "E. Pancini", Università di Napoli Federico II, Complesso Monte Sant'Angelo, Via Cinthia, I-80126 Napoli, Italy

[7] CNR-SPIN, Complesso Monte Sant'Angelo, Via Cinthia, I-80126 Napoli, Italy

[8] CNR-SPIN, Dipartimento di Fisica, Politecnico di Milano, Piazza Leonardo da Vinci 32, I-20133 Milano, Italy

[†] Present address: Department of Physics and Seitz Materials Research Laboratory, University of Illinois, IL-61801 Urbana, USA

[*] Corresponding author. E-mail: giacomo.ghiringhelli@polimi.it

**Abstract:** Charge density waves are a common occurrence in all families of high critical temperature superconducting cuprates. Although consistently observed in the underdoped region of the phase diagram and at relatively low temperatures, it is still unclear to what extent they influence the unusual properties of these systems. Using resonant x-ray scattering we carefully determined the temperature dependence of charge density modulations in $(Y,Nd)Ba_2Cu_3O_{7-\delta}$ for three doping levels. We discovered short-range dynamical charge density fluctuations besides the previously known quasi-critical charge density waves. They persist up to well above the pseudogap temperature $T^*$, are characterized by energies of few meV and pervade a large area of the phase diagram, so that they can play a key role in shaping the peculiar normal-state properties of cuprates.

**Main text**: High-$T_c$ superconductors (HTS) are doped Mott insulators, where the quasi-two-dimensionality of the layered structure and the large electron-electron repulsion (responsible, e.g., for the robust short-range antiferromagnetic correlations) make them deviating from the Landau Fermi liquid paradigm. The doping-temperature ($p$-$T$) phase diagram encompasses, at low $T$, the antiferromagnetic and the superconducting orders and, at higher $T$, the pseudogap region, which marks, below the cross-over temperature $T^*$, a reduction of the quasiparticle density of states in some sections of the Fermi surface. In the pseudogap state and up to optimal doping $p$~0.17, short/medium range incommensurate charge density waves (CDW) emerge as an order weakly competing with superconductivity. CDW were proposed theoretically since the early times of the high temperature superconductivity age (*1*,*2*,*3*); experimental evidence by surface and bulk sensitive techniques came initially in selected materials (*4*,*5*,*6*,*7*), and later in all cuprate families (*8*,*9*,*10*,*11*,*12*). Moreover long-range tridimensional CDW (3D-CDW) order has been observed inside the superconductive dome (for $p$~0.08-0.17) in special circumstances, e.g. in high magnetic fields that weaken superconductivity or in epitaxially grown samples (*13*,*14*,*15*). Finally, it has come as a surprise the recent observation of CDW modulations in overdoped $(Bi,Pb)_{2.12}Sr_{1.88}CuO_{6+\delta}$ outside the pseudogap regime too (*16*), hinting at a wider than expected occurrence of this phenomenon.

The relevance of charge density modulations for the unconventional normal state and the superconducting properties of HTS is currently debated. In some theoretical models long and short range CDW are seen as epiphenomena on top of a fundamentally peculiar metallic state, where the end-point at $T=0$ of the pseudogap boundary line ($p^*\sim 0.19$-$0.21$) marks the physical onset of a novel non-Fermi-liquid metallic phase (*17,18,19,20,21,22*). In an alternative scenario, charge density modulations are instead pivotal to the anomalous properties of cuprates (*1,23,24*). There CDW are expected to be critical, i.e. associated to the divergence of a correlation length at a quantum critical point QCP, and to permeate, by charge density fluctuations (CDF), a much broader area of the phase diagram. To establish to what extent static and fluctuating charge density waves contribute to the phase diagram, we have measured them in (Y,Nd)Ba$_2$Cu$_3$O$_{7-\delta}$ as a function of doping and temperature. We have discovered that CDF are present over a broad region of the phase diagram, which strengthen the importance of charge density modulations in determining the normal state properties of cuprates, and that the previously known short/medium range CDW appear as precursors of the long-range charge modulation detected in the presence of high magnetic fields, pointing towards a quasi-critical phenomenon.

We measured resonant inelastic x-ray scattering (RIXS) on four (Y,Nd)Ba$_2$Cu$_3$O$_{7-\delta}$ (YBCO and NBCO) thin films spanning a broad range of oxygen doping, going from the antiferromagnetic (AF) region, where $T^*$ is not even defined, passing through the underdoped (UD) regime, up to the optimally doped (OP) region (see Materials and Methods, and Supplemental Material Fig. S1) (*25,26*). Significantly, the latter case allowed us to cross the pseudogap line marked by the temperature $T^*$. Measurements have been performed at the Cu $L_3$ edge (~930 eV), over broad in-plane wave vector ($q_{//}$ = 0.2–0.4 reciprocal lattice units, r.l.u.) and temperature ranges ($T$ = 20K-270K). Figure 1C shows the quasi-elastic (near-zero energy loss) component of the RIXS spectra as a function of $q_{//}$ = ($H$,0) taken on sample UD60 (NBCO, $p\approx 0.11$) at different temperatures. A clear peak is present in the whole temperature range under investigation. The intensity of the peak decreases as the temperature increases, while it is little temperature dependent above 180-200K. A quasi-elastic peak, robust vs temperature, is also present on the samples UD81 (YBCO, $p\approx 0.14$, see Fig. 1B) and OP90 (NBCO, $p\approx 0.17$, see Fig. 1A). On the contrary, the antiferromagnetic sample shows no peaks above the linear background (see Fig. 1D). These data highlight the existence of a genuine quasi-$T$-independent scattering signal representative of a short-range charge order; hints of the presence of this peak were visible in resonant x-ray scattering data published previously but had never been discussed (*10*). It must be noted that no peak is present in the ($H$,$H$) direction, whose featureless linear shape can be used as indicative linear background in the fitting of the scans along ($H$,0) (see Figs. 2A-2C). Interestingly the scattering intensity is approximatively linear vs $1/T$ as shown in the inset of Fig. 1C: the extrapolation to very high temperature ($1/T$=0) provides an estimate of the intrinsic background of the signal, due mainly to scattering from low energy phonons and surface imperfections (see Supplemental Material, and Fig. S3).

We have decomposed the ($H$,0) scans by least square fitting to extract the peak intensity, width and position. Figure 2 shows results on sample UD60. At high temperatures, the quasi-elastic intensity can be fitted by assuming a single, broad Lorentzian profile on top of a linear background (see Fig. 2B). At lower temperatures, two peaks are necessary: a broad peak (BP), very similar to that measured at higher temperature, and a narrow peak (NP) centered at a nearby value of $H$. We have also scanned along $K$ while fixing $H=H_{NP}$ at the maximum of the NP in the ($H$,0) scan: there, again, the shape consists of a narrow and a broad peak, both centered at $K$=0. Since the temperature dependence of the $K$-scans follows that of $H$-scans (see Supplemental Material, Fig. S2), the quasi-elastic peak in the reciprocal space can be modeled by a double 2D Lorentzian, a broad one and a narrow one, centered respectively at $q_c^{NP}$ = (0.325, 0) and at $q_c^{BP}$ = (0.295, 0) (see Fig. 2E).

Figure 3 summarizes the outcome of the fittings for the samples UD60 and OP90, whereas Fig. S6 in the Supplemental Material reports the corresponding results for the UD81 sample. The NP presents all the characteristics previously observed in several underdoped cuprates and commonly attributed to the incommensurate CDW. The BP shares with the NP the position in the reciprocal space (although with small differences, see Fig. S7), but it has a very different, almost constant, temperature dependence. Therefore, we attribute the BP to very short-range charge modulations, i.e. to charge density fluctuations, as depicted by the reddish region of the $T$-$p$ phase diagram of Figure 4A. Noticeably, whereas the FWHM of the NP follows a critical temperature dependence, the temperature dependence of the BP width is much weaker in the accessible temperature range and within our experimental uncertainties. Finally, it is important to note that, although the amplitude of the NP (i.e. the peak height) is larger than that of the BP at low temperature, the total "volume" (i.e. the integrated scattering intensity) is always dominated by the BP, at least in the accessible $T$ range above $T_c$ (see Fig. 3).

To further clarify the double regime of the phenomenon and to assess the possible dynamical character of the CDF, we have studied the energy associated to the BP by exploiting the high resolution of our instrumentation. We measured Cu $L_3$ RIXS spectra on the OP90 and UD60 samples at selected temperatures and at the wave vector of the BP maximum. At all temperatures, the main peak is slightly broader than the instrumental resolution (40 meV) and is not centered at zero energy loss, with a stronger inelastic component at higher $T$ (see Supplemental Material). Contributions to this quasi-elastic peak are coming from phonons, from elastic diffuse scattering from the sample surface and from charge fluctuations. The phonon peak intensity is either $T$-independent (for those higher than 30 meV) or decreases upon cooling down (for those at lower energy); diffuse scattering is constant with $T$. The scattering related to charge density modulations is the only one expected to grow in intensity when decreasing $T$. Figure 4B shows the quasi-elastic component of three spectra taken on the optimally doped sample at 90K, 150K, and 250K, and $q_{//}$ = (0.31,0), after subtraction of the phonon contribution, as estimated from the ($H,H$) scan (details on how the additional phonon contribution and elastic scattering have been subtracted can be found in the Supplemental Material). To better single out the charge density contribution we have subtracted the higher $T$ spectra from the lower $T$ ones: in Figures 4C and 4D we show the 150K-250K and 90K-150K difference spectra. The resulting peaks are narrower than the original spectra. Moreover, the higher-$T$ case is evidently centered at $\omega_0$ ~15 meV, whereas the lower-$T$ difference is almost elastic. This means that the BP, still dominant at high $T$, has a fluctuating nature, whereas the NP emerges at lower $T$ as a nearly static, quasi-critical CDW.

These results can be interpreted within the theory mentioned in the introduction, based on the charge density instability of the high-doping correlated Fermi-liquid (*1,23,24*). We have fitted the three quasi-elastic peaks in Figure 4B (see also Fig. S11) by using the dynamical charge susceptibility, proportional to the correlation function of the density fluctuations

$\langle n(q,\omega)n(-q,\omega)\rangle$ in Fourier space, as obtained from a standard dynamical Ginzburg-Landau approach in Gaussian approximation (*27,28*). Its imaginary part, multiplied by the Bose function, gives the response function for the charge density modulation and represents the intensity of the low-energy peak in the RIXS spectra $I(q,\omega)$. Then the spectra in Fig. 4B can be fitted with this theory assuming a characteristic energy $\omega_0 \approx 15$ meV for these overdamped charge fluctuations at 150K and 250K; the 90K curve is better fitted with $\omega_0 \approx 7$ meV, indicating that the NP is associated to lower or null energy. Similar fittings for the UD60 sample give $\omega_0 \approx 5.5$ meV at high temperatures and 3.3 meV at 90K (see Supplemental Material for technical details on theory).

The broad peak is therefore generated by dynamical CDF, with pure 2D character related to individual CuO$_2$ planes, and characterized by a non-critical behavior. Its ultra-short-range nature is confirmed by a correlation length – given by the FWHM $\propto \xi^{-1}$ values - of $2\xi \approx 4a$, which is comparable to the modulation period (3.4$a$, see Supplemental Material, Fig. S8). The narrow peak, instead, comes from quasi-critical CDW appearing only below $T_{QC}$ (see crosses in the phase diagram of Figure 4A, and Supplemental Material). Quasi-critical CDW compete with superconductivity, as highlighted by the intensity and $\xi$ saturation (or decrease) below $T_c$ (see Figs. 3-S4-S5-S6). In the relatively narrow temperature range above the occurrence of such competition the linear extrapolation to zero of the NP width provides an estimate of the critical temperature $T_{3D}$, below which, in the absence of superconductivity, $\xi$ diverges and the CDW form a static 3D order. The values of $T_{3D}$ of our three samples (see Figs. 3C, 3D, S6C), indicated as squares in the phase diagram of Figure 4A, are in fairly good agreement with the onset of the long-range 3D CDW obtained in high magnetic fields by NMR and RXS experiments (*13,14*) (see blue region in Fig. 4A). For OP90, $T_{3D}$ is relatively low because the doping corresponds roughly to $p_c$ of the QCP. Moreover, a scan taken at 62K on overdoped YBCO (OD83, $p \approx 0.18$) shows that only the dynamical CDF survive at $p > p_c$ (see Supplemental Material, Fig. S9). Therefore, were it not for the competing superconducting order that quenches the critical behavior of CDW, 3D CDW would occur for $p < p_c$ and $T < T_{3D}$ without any application of magnetic field. The static 3D CDW dome is centered at $p \sim 1/8$ and is delimited by two QCPs at $p \approx 0.08$ and 0.17 (*29,30*); inside that doping range, above $T_{3D}$ and below $T_{QC}$, quasi-critical CDW, precursors of the static 3D CDW, are present.

The phase diagram of Fig. 4A visualizes the scenario of a continuous crossover from the pure 2D dynamical CDF at high $T$ and all dopings, to a quasi-critical CDW, still 2D, below $T_{QC}$ and for 0.08<$p$<0.17, to the static 3D CDW usually hindered by superconductivity. Although disregarded up to now, dynamical CDF represent the bulk of the iceberg of the CDW phenomenon in cuprates. In fact, they pervade a large part of the phase diagram and coexist with both quasi-critical CDW and, possibly, 3D CDW (see Fig. 4A) and their total scattering intensity (the volume of the associated BP) dominates at all $T$ (see Fig. 3E,F). Moreover, they do not compete with superconductivity.

This picture is consistent with the old theoretical proposal by Castellani *et al.* (*1*) pivoting around the QCP at $p_c$. Due to the weak coupling of CuO$_2$ planes, CDW have a marked 2D character and, due to strong quantum-thermal dynamical fluctuations, they acquire a truly static character only below $T_{3D}$, which is, for (Y,Nd)BCO, smaller than $T_c$, thus requiring strong magnetic fields or epitaxially-grown samples to suppress superconductivity to obtain static 3D CDW.

Although this theory can explain most of the experimental findings, some questions arise. First, other cuprate families will have to be tested and the doping region extended, in order to confirm the general applicability of the dynamic CDF scenario. A BP, centered at $q_{//} \approx q_c^{NP}$ and persisting at high temperatures, has been indeed observed recently in other cuprates, both in the underdoped (*31,32*) and in the overdoped (*33*) regime (in the latter case accompanied by an anomaly in the lattice vibrations), pointing toward a universality of the CDF phenomenon. However, none of the aforementioned experiments has been conclusive in this respect, since a complete temperature dependence and/or a discrimination of the quasi-elastic signal from the inelastic one was still missing. The actual relation between the quasi-critical CDW and the dynamical CDF has to be fully clarified too, with particular reference to the possible spatial separation or coexistence of the two phenomena, ultimately linked to the role of disorder in the samples studied by STM experiments (*7,34,35*) and micro-XRS (*36*).

The most intriguing finding of this work is the ubiquitous presence (both in temperature and doping) of a broad peak due to dynamical charge density fluctuations with two specific relevant features: they have small energies of a few meV and still they extend over a broad range in momentum space. Therefore, they provide a quite effective low-energy scattering mechanism for all the quasiparticles on the Fermi surface. This makes these excitations an appealing candidate for producing the linear temperature dependence of the resistivity in the normal state and the other marginal Fermi liquid phenomena that, since the early times (*37*), constitute the peculiar properties of HTS cuprates.

**Acknowledgments:**

This work was supported by ERC-P-ReXS project (2016-0790) of the Fondazione CARIPLO and Regione Lombardia, in Italy. R. A. is supported by the Swedish Research Council (VR) under the project "Evolution of nanoscale charge order in superconducting YBCO nanostructures". The authors acknowledge insightful discussions with B. Keimer, M. Le Tacon, T. P. Devereaux, M. Moretti, M. Rossi, W.S. Lee. The experimental data were collected at the beam line ID32 of the European Synchrotron (ESRF) in Grenoble (France) using the ERIXS spectrometer designed jointly by the ESRF and Politecnico di Milano.

**Author contributions:**

G.G., M.G., L.B., C.D.C. and R.A. conceived and designed the experiments with suggestions from S.C., N.B.B. and M.S.; R.F., Y.Y.P., G.G., L.B., R.A., G.D.V., G.M.D.L., M.S., E.A., F.L., D.B., and N.B.B. performed the RIXS measurements. R.A., G.G. G.D.V., Y.Y.P. and L.B. analysed the RIXS experimental data; S.C., C.D.C. and M.G. performed the theoretical calculations. G.M.D.L. and M.S. grew and characterized the NBCO films; R.A., E.A. and F.L. grew and characterized the YBCO films. R.A., G.G., M.G., and C.D.C. wrote the manuscript with the input from F.L., L.B., N.B.B., Y.Y.P., and contributions from all authors.

**Competing financial interests:**

The authors declare no competing financial interests.

**Figures:**

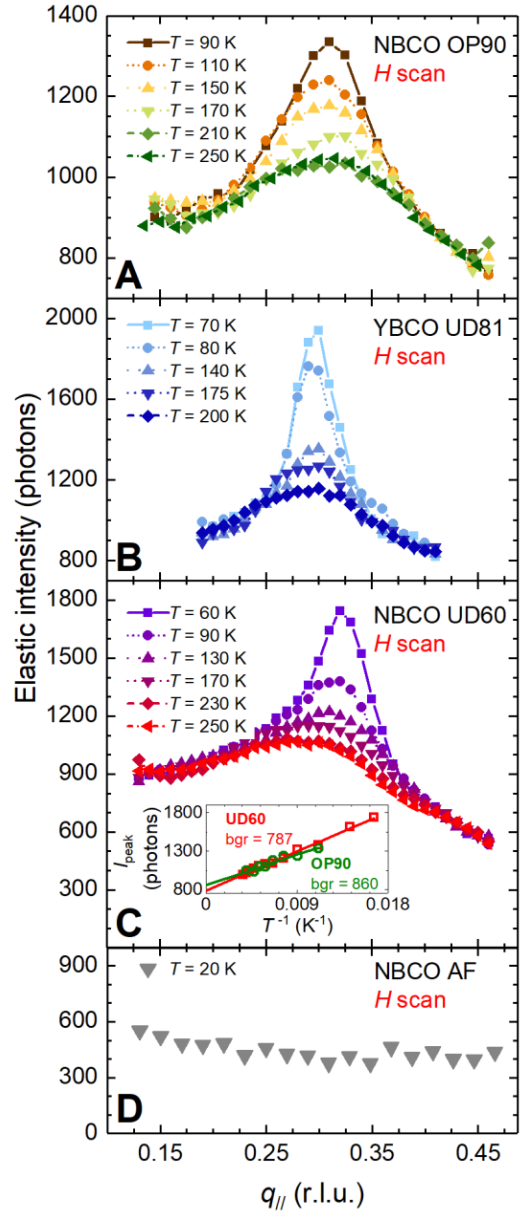

**Fig. 1**: **Quasi-elastic scan along the ($H$,0) direction for several (Y, Nd)Ba$_2$Cu$_3$O$_{7-\delta}$ films with different oxygen doping.** The quasi-elastic intensity has been determined by integrating the Cu $L_3$ RIXS spectra measured at different $q_{//}$ values in the energy interval [-0.2 eV, +0.15 eV]. The measurements have been performed at different temperatures on the following samples: **(A)** Optimally doped NBCO, $p \approx 0.17$. **(B)** Underdoped YBCO, $p \approx 0.14$. **(C)** Underdoped NBCO, $p \approx 0.11$. **(D)** Insulating NBCO, $p < 0.05$. The inset in panel **(C)** shows the peak intensity $I_{peak}$ vs $T^{-1}$ for samples OP90 (circles) and UD60 (squares). The extrapolation to $T \to \infty$ provides an estimate of the intrinsic background of the signal.

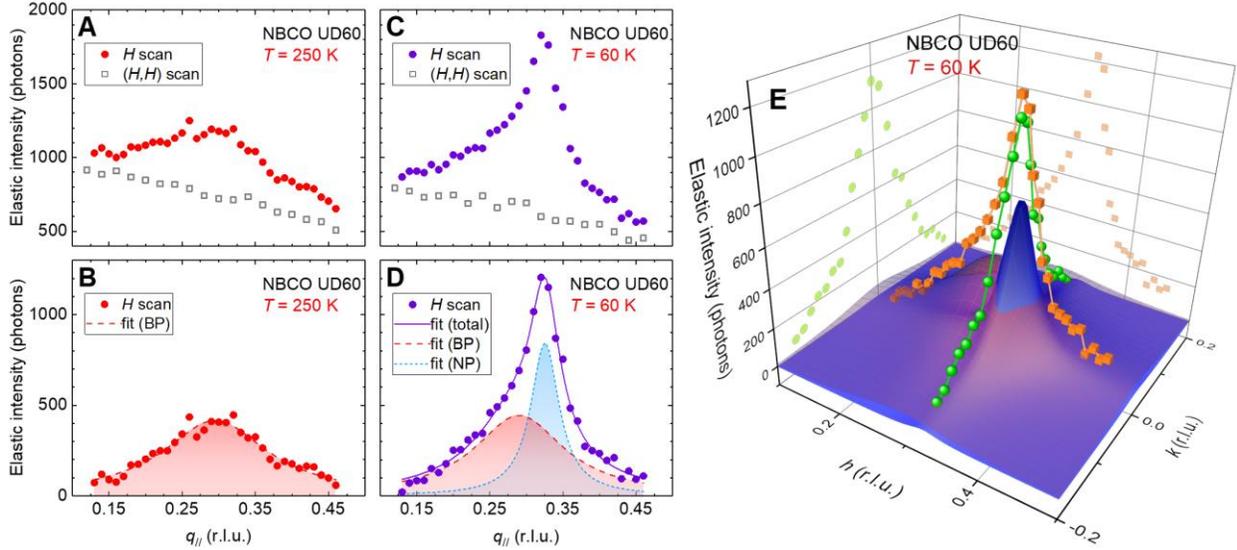

**Fig. 2**: **Two distinct peaks. (A)** Quasi-elastic scan measured along ($H$,0) on sample UD60 at $T$ = 250 K (red circles). **(B)** After subtracting the linear background, given by the quasi-elastic scan measured along the Brillouin zone diagonal (open squares in panel A), a clear peak is still present, which can be fitted by a Lorentzian profile (dashed line). **(C)** Same as A, but at $T$ = 60 K (violet circles). **(D)** After subtracting the linear background (open squares in panel C), the data can be fitted assuming the sum of two Lorentzian profiles (solid line), one broader (dashed line), similar to that measured at 250 K, and the second one narrower and more intense (dotted line). **(E)** The 3D sketch shows the quasi-elastic scans measured along $H$ (cubes) and along $K$ (spheres) at $T$=60 K on sample UD60, together with the Lorentzian profiles used to fit them. A narrow peak (NP, blue surface) emerges at $q_c^{NP}$=(0.325, 0) from a much broader peak (BP, red surface) centered at $q_c^{BP}$=(0.295, 0).

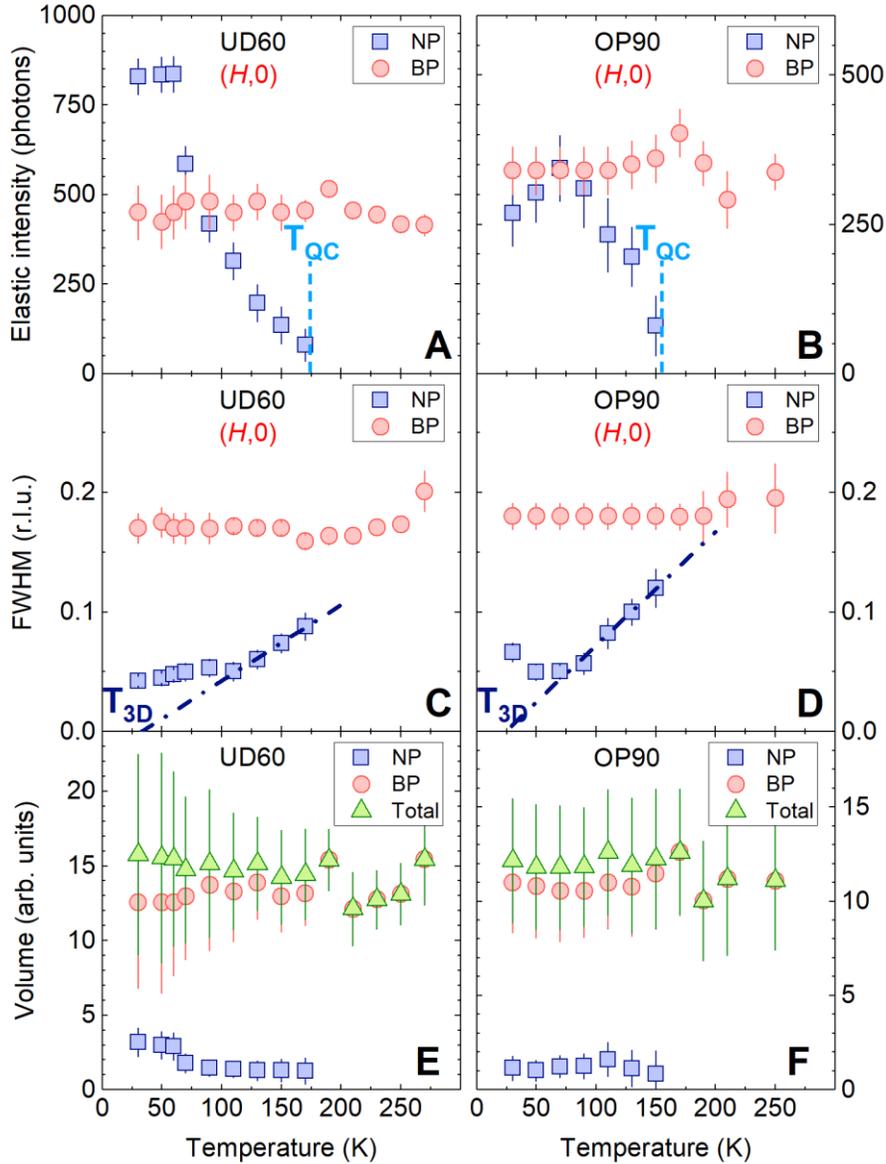

**Fig. 3**: **Characteristics of the two charge density modulation peaks.** Temperature dependence of the parameters of the two Lorentzian profiles, used to describe the quasi-elastic peaks of samples UD60 and OP90 (squares refer to the narrow peak, circles to the broad peak). **(A-B)** Intensity. **(C-D)** FWHM. $T_{QC}$ is 175 K for sample UD60 and 155 K for sample OP90. $T_{3D}$ is 33 K for sample UD60 and 24 K for sample OP90. **(E-F)** Volume of the charge density modulations. The total volume (triangles), given by the sum of the volumes of the two peaks, is dominated by the broad peak.

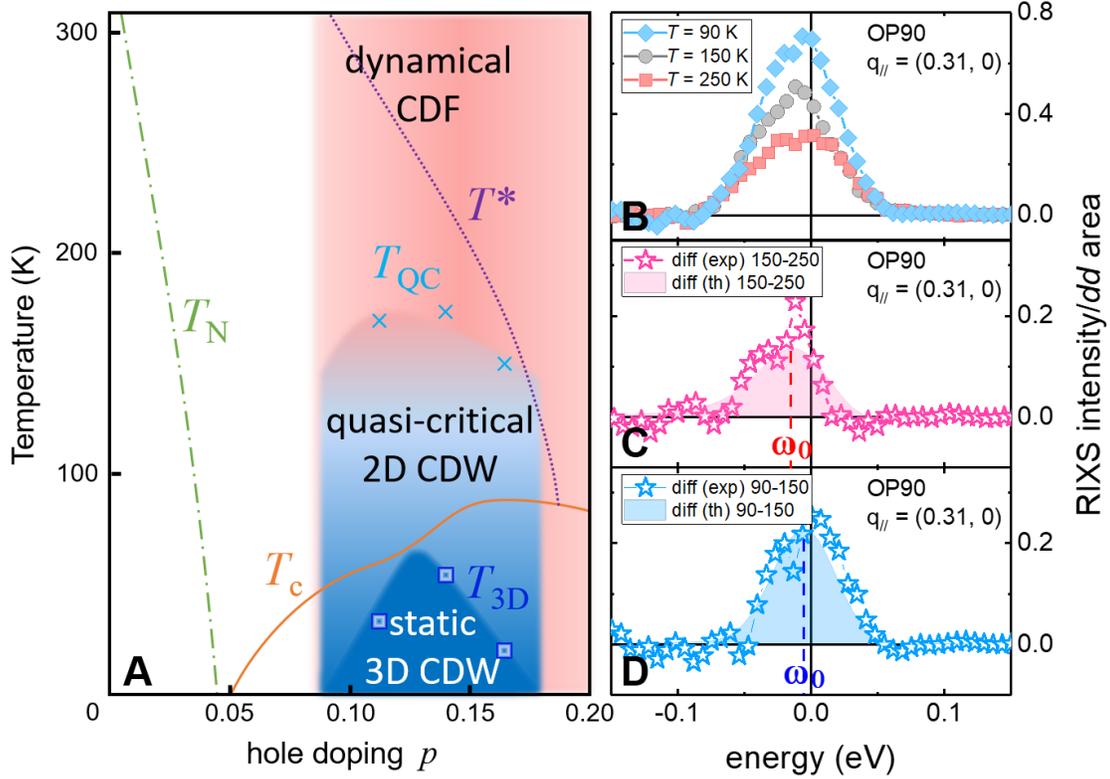

**Fig. 4**: **Static and dynamic charge order in the phase diagram of the HTS cuprates. (A)** The *T-p* phase diagram of cuprates is typically marked by the antiferromagnetic, the pseudogap and the superconducting regions (each characterized by the onset temperatures $T_N$, $T^*$ and $T_c$). Our results prove that most of these regions are pervaded by charge density modulations of some sort. The narrow peak describes the CDWs, manifesting in a region (pale blue) below $T_{QC}$ (crosses). These 2D CDWs are quasi-critical, and precursors of the static 3D-CDW (blue region). Even though we cannot directly access this dome without a magnetic field, the temperatures $T_{3D}$ (squares) we have inferred from the *T* dependence of the NP FWHM, are in agreement with those, previously determined in NMR and RXS experiments (*13,14*). The broad peak describes short range charge density fluctuations (CDF), which dominate the phase diagram (red region), coexisting both with the quasi-critical 2D-CDW and with superconductivity, and persisting even above $T^*$. On the contrary, CDF disappear in undoped/antiferromagnetic samples (white region), while their occurrence between $p$~0.05 and $p$~0.08 has still to be determined. To evaluate the characteristic energies $\omega_0$ associated to the BP, we have measured high resolution RIXS spectra at various temperatures on the samples OP90 and UD60. **(B)** Quasi-elastic component of the spectra (after subtraction of the phonon contribution) at *T* = 90, 150 and 250 K, measured on sample OP90 at $q_{//}$ = (0.31, 0). **(C)-(D)** The experimental differences 150 K-250 K and 90 K-150 K spectra, presented in panel (B), are shown (stars), together with the theoretical calculation (solid areas). The data are in agreement with the theory assuming $\omega_0 \approx$ 15 meV at both 150 K and 250 K, and $\omega_0 \approx$ 7 meV at 90 K.

# SUPPLEMENTAL MATERIAL for

# Dynamical charge density fluctuations pervading the phase diagram of a Cu-based high-$T_c$ superconductor

R. Arpaia, S. Caprara, R. Fumagalli, G. De Vecchi, Y.Y. Peng, E. Andersson, D. Betto, G. M. De Luca, N. B. Brookes, F. Lombardi, M. Salluzzo, L. Braicovich, C. Di Castro, M. Grilli, G. Ghiringhelli[*]

* Corresponding author. E-mail: giacomo.ghiringhelli@polimi.it

**Supplementary Materials include:**

Materials and Methods

Supplementary text

Figures S1 – S11

References (38 – 46)

**Materials and Methods**

*Sample characterization*

YBa$_2$Cu$_3$O$_{7-\delta}$ (YBCO) films, having a thickness of 50 nm, have been deposited by pulsed laser ablation on 5 × 5 mm$^2$ MgO (110) substrates. For the experiment we have used an underdoped film, UD81 ($p \approx 0.14$), and a slightly overdoped film, OD85 ($p \approx 0.18$). Details on the sample growth and characterization have been previously reported (*26*). An in-situ annealing at 790 °C has been done before the deposition to induce a surface reconstruction of the substrate, favoring the growth of untwinned films. After the deposition, the UD81 and the OD85 have been slowly cooled down at an oxygen pressure respectively of 0.7 μbar and 900 mbar. For both the films the untwinning degree, determined via x-ray diffraction (XRD) 2$\theta$-$\omega$ measurements of the (038)/(308) reflections, exceeds 80%.

Nd$_{1+x}$Ba$_{2-x}$Cu$_3$O$_{7-\delta}$ (NBCO) films, having a thickness of 100 nm, have been deposited by high-oxygen-pressure diode sputtering (heater temperature ≈ 900 °C, oxygen pressure ≈ 2 torr) on 10 × 10 mm$^2$ SrTiO$_3$ (001) substrates (*25*). For the experiment we have used an antiferromagnetic film, AF (x = 0, $p < 0.05$), an underdoped film, UD60 (x = 0.2, $p \approx 0.11$), and an optimally doped film, OP90 (x = 0, $p \approx 0.17$). The AF film is achieved by in-situ annealing of an optimally doped NdBa$_2$Cu$_3$O$_7$ film in argon atmosphere (at a pressure of 10 mbar for 48 h). The UD60 NBCO film is untwinned, while the OP90 sample is twinned.

The critical temperature $T_c$ and the pseudogap temperature $T^*$ have been determined by measuring the resistance vs temperature $R(T)$ of the films with a 4-point setup (see Supplemental Material, Figure S1). The doping $p$ for each film has been determined by the knowledge of the $T_c$ values, in combination with the c-axis length obtained via XRD (see Refs. *26*, *38* for more details). For the NBCO films the $T_c$ vs doping dependence is in excellent agreement with known relationship between Seebeck effect and carrier density, as shown in Ref. (*25*). An in-vacuum preparation of the surface of the films before the RIXS investigation is not needed.

*RIXS measurements*

To properly investigate the possible presence of charge density fluctuations (CDF) at high temperature, the quality of the RIXS experimental setup represents a crucial factor. Measurements at high temperature have been performed in all the experiments, mostly based on Resonant Elastic X-ray scattering (REXS), focusing on charge density wave (CDW) in recent years. Yet, no evidence of charge order close to room temperature has been shown in the UD regime. The reason has to be found in the procedure commonly used to reveal the presence of the CDW order component: The REXS scans at different temperatures are subtracted by a reference scan measured at a threshold temperature, above which the scans are unchanged ($\approx$ 180-200 K in case of YBa$_2$Cu$_3$O$_{7-\delta}$). This is due to the large background affecting the REXS scans, connected to the inelastic component of the spectra which, differently than in RIXS, cannot be separated from the elastic/quasi elastic component (whose changes are directly connected with the presence of a charge order). This procedure intrinsically hides the presence of any charge orders above the threshold temperature.

The RIXS measurements have been performed at the beamline ID32 of the European Synchrotron Radiation Facility (ESRF, Grenoble, France) using the high-resolution ERIXS spectrometer (*39*). The samples were mounted on the 6-axis in-vacuum Huber diffractometer/manipulator. The resonant conditions were achieved by tuning the energy of the incident X-ray to the maximum of the Cu $L_3$ absorption peak, around 931 eV. The instrumental energy resolution was set at 40 meV, determined as the full width at half maximum of the non-resonant diffuse scattering from the silver paint. The RIXS experimental geometry has been shown in previous papers (*16,40*): X-rays are incident on the sample surface and scattered at an angle 2θ. Momentum transfers are given in reciprocal lattice units, i.e. in units of the reciprocal lattice vectors $a^* = 2\pi/a$, $b^* = 2\pi/b$, $c^* = 2\pi/c$. These units are therefore defined by using the tetragonal unit cell with $a = b = 3.905$ Å and $c = 11.75$ Å for NBCO films and the orthorhombic unit cell with $a = 3.82$ Å, $b = 3.88$ Å and $c = 11.76$ Å for YBCO films. For both NBCO and YBCO films the *c*-axis is normal to the sample surface. The sample can be rotated azimuthally around the *c*-axis to choose the in-plane wave vector component. The measurements presented in the paper were taken with 2θ = 149.5°, giving $|q| = 0.91$ Å$^{-1}$, which allows one to cover the whole first Brillouin zone along the [100] direction (0.5 r.l.u. $\approx$ 0.81 Å$^{-1}$). The total acquisition time for each RIXS spectrum was 2 minutes (sum of individual spectra of 30 seconds) for the $q_{//}$ and temperature dependence measurements (see Figs. 1-2) and 120 minutes (sum of individual spectra of 2 minutes) for the high-resolution measurements (see Fig. 4). The exact position of the elastic (zero energy loss) line has been determined by measuring, for each $q$, a non-resonant spectrum of silver paint. In addition to that, for the high-resolution spectra (see Fig. 4) the position of the elastic line has been confirmed by measuring a resonant spectrum at specular reflection. The quasi-elastic intensity was determined by the integral in the range [-0.2 eV, +0.15 eV]. In Fig. 4 the RIXS spectra have been normalized to the integral of the inter-orbital transitions (*dd* excitations) in the range [-3 eV, -1 eV]. For the twinned NBCO films, only the CDW peak along the [100] direction has been explored: The *K* scan (as in Fig. S2) is performed at a value of *H* coinciding with the position of the maximum of the narrow peak, $H_{NP}$.

*Characteristics of the broad peak*

The first question arising, when investigating the quasi elastic component of the RIXS spectra at high temperature, is whether such temperature independent broad peak (BP) is just a background signal or it has the same nature as the temperature-dependent peak, measured at lower temperatures, which is commonly associated in literature to the presence of the CDW order (the narrow peak, NP).

First of all, the quasi-elastic signal of the BP has a maximum at a finite $q_{//} = (H, 0)$ and, out of the Cu $L_3$ resonance, it completely disappears at any measured doping, even at high temperatures.

The BP emerges only in the samples where superconductivity is also present, while in the antiferromagnetic sample only a linear dependence with $q_{//}$ has been measured at any temperature (see Fig. 1D).

The BP, at any temperature, is present only when measuring the RIXS spectra along the in-plane crystallographic axes of the $(Y,Nd)Ba_2Cu_3O_{7-\delta}$ unit cell, i.e. along the Cu-O bond directions $(H, 0)$ and $(0, K)$.

The BP is, although weakly, temperature dependent. To make this dependence more visible it is useful to plot the temperature dependence of the total scattering intensity $I_{peak}$ (that is without distinguishing the BP and NP contributions) in order to see whether the clear *T*-dependence of the NP (that clearly reflects in the total $I_{peak}$) also continues on the total scattering intensity at temperatures *T*>170K, where only the BP is present. As shown in the inset of Fig. 1C, the total scattering intensity $I_{peak}$, measured at the $q_{//}$ of the peak maximum, is approximatively linear vs $1/T$ in the whole temperature range under investigation. From the analysis of $I_{peak}$ vs $1/T$ we have determined, through the extrapolation of $I_{peak}$ at infinite

$T$, an estimate of the intrinsic background of the signal, which appears to be similar for the two NBCO samples under investigation (≈800 photons). The background level estimated through this procedure is in fairly good agreement with the value we have measured along the ($H,H$) direction, which we have therefore used as indicative background in the fitting of the scans along ($H,0$). The peak intensity $I_{peak}$, after subtraction of the background level, follows a power law with a critical exponent α = -1 for both samples (see Fig. S3).

The BP is not affected by the twinning state of the films: in samples UD60 and UD81, where aligned CuO chains are present throughout most of the samples, the quasi elastic signal vs $q_{//}$ presents the same behavior vs temperature (see Figs. 1B-1C) observed in sample OP90, which is instead twinned (see Figs. 1A).

A further confirmation of the common nature of broad and narrow peaks comes from the measurement of the FWHM of the peaks in the various samples under investigation. Indeed, the NP is narrower in the YBCO films than in the NBCO films (for instance, at $T_c$ the FWHM of the NP is ≈0.03 r.l.u. for the sample YBCO UD81, while it is ≈ 0.05 r.l.u. for the samples NBCO UD60 and OP90). Similar dependence of the FWHM vs twinning occurs also for the BP (the FWHM of the BP is ≈ 0.15 r.l.u. for the samples YBCO UD81 and OD83, while it is ≈ 0.18 r.l.u. for the samples NBCO UD60 and OP90).

Finally, the BP cannot be associated to intrinsic disorder of the samples, and therefore described in mere terms of a broadening in $q_{//}$ of the NP. In the latter case, indeed, the BP should be characterized by the same temperature dependence of the NP. On the contrary, its intensity and FWHM are characterized by peculiar temperature dependences, distinct from those of the NP (see next subsection for more details).

*Analysis on the temperature dependence of the fit parameters*

The NP, related to quasi-critical 2D-CDW, is characterized by an intensity and, simultaneously, a correlation length $\xi \propto \text{FWHM}^{-1}$, both increasing when the temperature decreases. Finally, when the temperature approaches the critical temperature $T_c$ of the film, both the intensity and the correlation length saturate (see Figs. 3, S4, S5, S6). This saturation is caused by the appearance of superconductivity, which is likely competing with the charge order. Significant temperatures for these quasi-static charge fluctuations are $T_{QC}$ and $T_{3D}$. The temperature $T_{QC}$ is the onset temperature, where the quasi-static order appears. The temperature $T_{3D}$ is the temperature where the quasi-static order would become static (FWHM = 0 → ξ = ∞) in absence of superconductivity. For each sample, both these temperatures can be determined from the intensity and FWHM vs $T$ behaviors of the NP. The $T_{QC}$ has been inferred from the intensity vs $T$ behavior as the temperature where the number of photon counts equals the uncertainty of the experimental data points (≈ 40 photons). In order to determine $T_{3D}$, we have fitted the FWHM vs $T$, in the temperature range above the occurrence of superconductivity and of superconducting fluctuations (*26,41*). The $T_{3D}$ is obtained from the extrapolation to FWHM = 0 of the linear dependence in temperature of the NP width.

At each temperature, the volume of the charge order peak is given by the product of the full width at half maximum of the quasi elastic peak measured along $H$ and along $K$, and the corresponding (geometrical) average intensity: $V = FWHM_H \cdot FWHM_K \cdot \sqrt{I_H \cdot I_K}$. The error on the volume, due to the errors $\sigma_{FWHM}$ and $\sigma_I$, relative to the FWHM and to the intensity respectively, is given by: $\sigma_V = \left( \frac{\sigma_{FWHM,H}}{FWHM,H} + \frac{\sigma_{FWHM,K}}{FWHM,K} + \frac{1}{2}\frac{\sigma_{I,H}}{I,H} + \frac{1}{2}\frac{\sigma_{I,K}}{I,K} \right) \cdot V$.

**Supplementary text**

*Analysis of the high resolution RIXS spectra: The CDW instability*

The proposal of a CDW instability in cuprates (or a close proximity to it due to low dimensionality, disorder and the competing superconducting phase) is quite old (*1,2,23,42*) and it is based on the mechanism of frustrated phase separation. In this framework, strong correlations tend to weaken the metallic character of the system giving rise to a Fermi liquid with a substantial enhancement of the quasiparticle (QP) mass m* and a sizable residual interaction. In this situation, weak or moderate additional attractions due to phonons or short-range magnetic interactions provide an effective mechanism that would induce an electronic phase separation with charge segregation on large scales. Since this is forbidden by the long-range Coulombic repulsion, the system chooses a compromise segregating charge on shorter scale, while keeping charge neutrality at large distance. The simplest outcome is then a CDW state, which tends to be formed where correlations are more important, i.e. by underdoping the system. Notice that the CDW wavelength and direction are not ruled by a Fermi surface instability (nesting, van Hove singularities and so on may play a secondary role), but are largely due to the energetic balance between the energy gain in charge segregation and Coulombic cost.

All this physics can be effectively cast in the simple language of many-body theory within a Strongly-Correlated-Random-Phase-Approximation (SC-RPA) considering a Fermi liquid with QP interacting via an effective interaction of the form $V(\mathbf{Q}) = U(\mathbf{Q}) - \lambda_\mathbf{Q} + V_c(\mathbf{Q})$.

$U(\mathbf{Q})$ is the QP residual repulsion stemming from the original bare large repulsion (1,2,24,42,43), $\lambda_\mathbf{Q}$ is an additional attraction, which may arise from phonons or non-critical spin interaction (like, e.g., a nearest-neighbour magnetic coupling), and $V_c(\mathbf{Q})$ is the Coulombic repulsion. In this framework, the effect of CDW with wavevector $\mathbf{Q}$ is described by the charge susceptibility, which in the SC-RPA can be written as

$$\chi(\mathbf{Q}, \omega) = \frac{\Pi(\mathbf{Q},\omega)}{1+V(\mathbf{Q})\Pi(\mathbf{Q},\omega)} \qquad (1)$$

Here $\Pi(\mathbf{Q}, \omega)$ is the Lindhard polarization function describing the mutual screening of the electrons. Its amplitude depends on the DOS at the Fermi level $N(E_F)$ (with $N(E_F)=\Pi(\mathbf{Q}=0,\omega=0)$, the larger their DOS, the better the electrons screen their mutual interactions). In turn the DOS is proportional to the electron effective mass $m^*$, $N(E_F)=m^*g(E_F)$, where $g(E_F)$ depends on the details of the band structure (e.g. the presence of a van Hove singularity). The effects of strong correlations are contained in the form of $V(\mathbf{Q})$ and in the QP effective mass entering $\Pi(\mathbf{Q}, \omega)$.

A charge instability at a given wave vector $\mathbf{Q}_{co}$ is obtained when the denominator in Eq. (1) vanishes leading to a diverging charge susceptibility, $[1 + V(\mathbf{Q}_{co})\Pi(\mathbf{Q}_{co}, \omega = 0)] = 0$.

The correlation function (propagator) of the charge fluctuations near the CDW instability is obtained by expanding both $\Pi(\mathbf{Q},\omega)$ and $V(\mathbf{Q})$ around $\mathbf{Q}_{c0}$ and $\omega=0$. The CDW collective modes are obtained by expanding the denominator of $\chi(\mathbf{Q}, \omega)$ for small frequency and small deviations from the instability wave vector $\mathbf{Q}_c$, obtaining the CDW correlation function (propagator)

$$D(\mathbf{Q}, \omega) = \frac{g^2}{\omega_0 + \nu(\mathbf{Q}) - i\omega - (\omega^2/\Omega)} \qquad (2)$$

It is worth noticing that this propagator could as well be obtained from the Ginzburg-Landau (GL) approach for a quantum CDW order parameter according to the standard procedure introduced by Hertz (*27,28*). In this latter framework, $\omega_0$ is the mass term of the GL functional, $v(\mathbf{Q}) \approx v|\mathbf{Q} - \mathbf{Q}_c|^2$ is the dispersion law of CDW fluctuations and it arises from the quadratic gradient term of the GL functional ($v$ is an electronic energy scale since we work with dimensionless momenta, measured in inverse lattice spacing $1/a$). The $i\omega$ term, stemming from the imaginary part of the Lindhard function gives rise to the Landau damping of the CDW fluctuations.

The poles of this propagator describe the dispersion of the nearly critical CDW modes. $g$ is the coupling between the fermionic QP and the CDW fluctuations. $\Omega$ is the energy scale above which the CDW critical mode becomes less overdamped and more propagating. The CDW characteristic energy scale $\omega_0 = v\xi^{-2}$ is the most relevant parameter to determine criticality. On the CDW instability line (if any) $\xi$ diverges and the CDW energy scale $\omega_0$ vanishes. Of course, in real materials the situation may be far more complicated. First of all, the CDW may be prevented by low dimensionality (in 2D a truly long-ranged incommensurate CDW is prohibited) and competing phases (superconductivity).

*Subtraction of the phonon contribution*

Figure S10A shows the quasi-elastic component of the high resolution (HR) RIXS spectra at $T$ = 150 and 250 K, measured on sample OP90 ($p \approx 0.17$) at $\boldsymbol{q}_{//}$ = (0.31, 0). At both the investigated temperatures, the main contribution to the charge order is given by the BP. The total peak intensity increases at the lower temperature, which is due to the presence of a weak contribution of quasi-critical 2D-CDW (see Fig. 3A-3B). The temperature variation of the quasi-elastic component has two main sources: charge order and phonons. Before fitting the peak, to be able to extract information about the charge order (i.e. the characteristic energy $\omega_0$) we have removed the temperature variation of the quasi-elastic component due to phonons. For this purpose, we have first measured at different temperatures high resolution RIXS spectra at $\boldsymbol{q}_{//}$ = (0.22, 0.22), where the contribution of any charge order is negligible (see for instance Fig. 2A-2C-2E) and any variation of the quasi-elastic peak can be attributed to phonons (see Fig. S10B). Then, at each temperature we have subtracted this spectrum from that measured at $\boldsymbol{q}_{//}$ = (0.31, 0) (see Fig. S10C). The difference spectrum

(squares in Fig. S10C, corresponding to the spectrum also shown in the main manuscript, circles in Fig. 4B), which represents the CDW component of the HR RIXS spectrum, is finally fitted within the CDW instability theory (solid area in Fig. S10C). It is worth mentioning that the difference spectrum we determine from these "phonon-removed" spectra is very similar to that we obtain considering the raw RIXS spectra (see Fig. S10D); in particular, from the fit we can determine the same characteristic energy $\omega_0 \approx 15$ meV. We can therefore conclude that the spectrum measured at $q_{//} = (0.22, 0.22)$ has no correlation with that measured at $q_{//} = (0.31, 0)$, and that the difference of the raw spectra shown in Figure S10A mainly originates from the dynamic short-range CDF.

*Doping and temperature dependence of the BP characteristic energy*

To study the evolution with doping and temperature of the characteristic energy $\omega_0$ of the BP, we have taken high resolution RIXS spectra also at $T = 90$ K, and on the sample UD60 ($p \approx 0.11$). Figure S11 shows a summary of all the performed measurements and fits. The result is that a drop of $\omega_0$ occurs both at $p \approx 0.11$, with respect to $p \approx 0.17$, and at 90 K, with respect to 150 and 250 K (the extracted $\omega_0$ values are listed in the caption of Figure S11). The lower $\omega_0$ values determined, at a fixed temperature, at $p \approx 0.11$ with respect to that at $p \approx 0.17$ agree with a scenario, which becomes more dynamic at doping approaching a QCP close to the OP. The drop of $\omega_0$ at 90 K, for both $p$ values, is in agreement with a quantum critical character of the dynamic charge density fluctuations. However, a possible role of the narrow peak in decreasing the $\omega_0$ value cannot be excluded. Indeed, in the fit only one peak is considered, to which one characteristic energy $\omega_0$ is associated. At 90 K, differently than at 150 and 250 K, also the narrow peak is robust and it might have a significant role in determining shape and position of the final spectrum. Since the two contributions (BP and NP) cannot be disentangled, the $\omega_0$ we have extracted from the fit at $T = 90$ K includes contribution from both dynamic CDF and quasi-critical CDW.

**Figures**

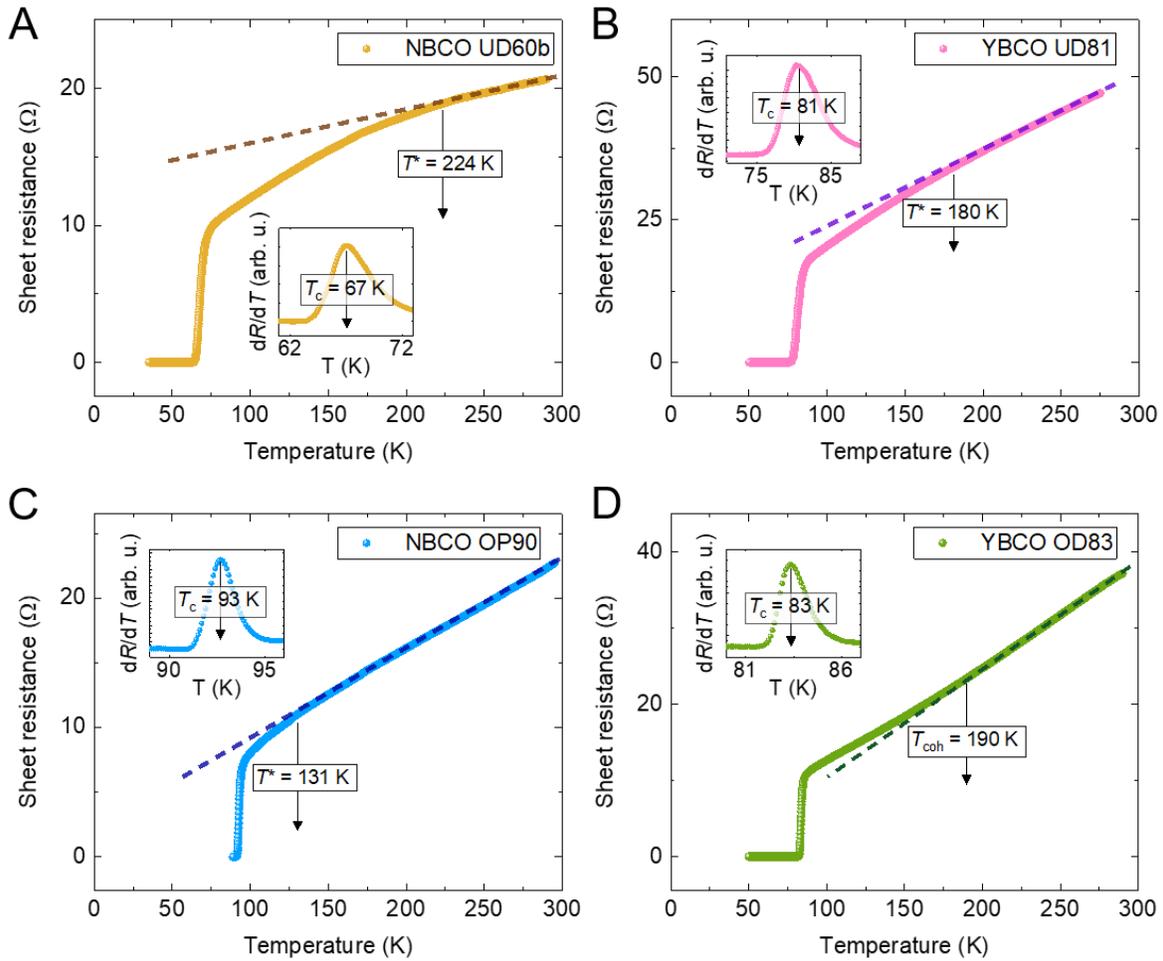

**Fig.S1: Sheet resistances of the (Nd,Y)BCO films as a function of the temperature.** The sheet resistance $R_\square$ is plotted as a function of the temperature for **(A)** sample UD60b (thickness: 100 nm; deposited in the same conditions as sample UD60); **(B)** sample UD81 (thickness: 50 nm); **(C)** sample OP90 (thickness: 100 nm); **(D)** sample OD83 (thickness: 50 nm). The critical temperature $T_c$ of the films has been extracted from the maximum of the first derivative of the $R(T)$ characteristic (inset in the four panels). The pseudogap temperature $T^*$ is instead inferred by the departure from the linear $R(T)$ behaviour at high temperature, which is a signature of the strange metal phase of cuprates (dashed line in the four panels). For the sample OD83, the temperature $T^*$ is not defined. Indeed the sample is slightly overdoped, as highlighted by the curvature of the $R(T)$ characteristic at lower temperature, which is opposite with respect to that characterizing the underdoped films. In this doping range, the temperature which can be inferred by the departure from the linear $R(T)$ behaviour at high temperature is the so-called coherence temperature $T_{coh}$ (*44,45*). $T_{coh}$ represents a crossover temperature from a coherent to an incoherent metal state, which has been observed by means of angle resolved photoemission spectroscopy (*46*).

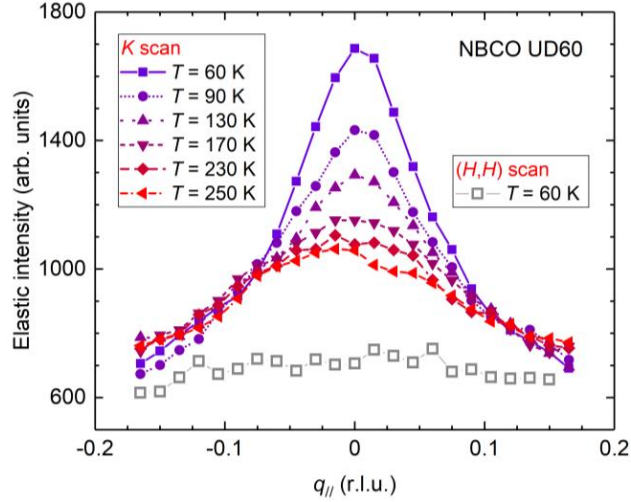

**Fig.S2: Quasi-elastic intensity measured at different temperatures along the $K$ direction on sample UD60.** Similarly to what we have measured along the $(H, 0)$ direction, a quasi-elastic peak (solid symbols) is present up to the highest investigated temperature ($T$ = 270 K) along the $(H_{NP}, K)$ direction. The peaks measured at temperatures higher than ≈ 170 K, whose intensity is only weakly temperature-dependent, are totally different from the background, measured along the Brillouin zone diagonal (open squares).

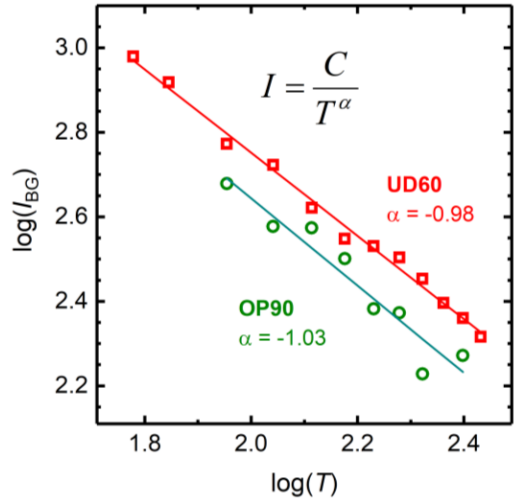

**Fig.S3: Dependence of the total scattered intensity with temperature, after subtraction of the background level.** The intensity $I_{BG}$, given by $I_{peak} - bgr$, where $I_{peak}$ is the scattering intensity, measured at the $q_{//}$ of the peak maximum, and $bgr$ is the background level estimated from the $I_{peak}(T^{-1})$, is plotted vs $T$ in a log-log graph, in order to get the critical exponent α, in the hypothesis of $I=C/T^{\alpha}$ (C being a coefficient of proportionality). Both the investigated NBCO samples are characterized by α ~ -1. This quasi critical behaviour motivates a more insightful analysis that takes into account the shape of the scattering profiles.

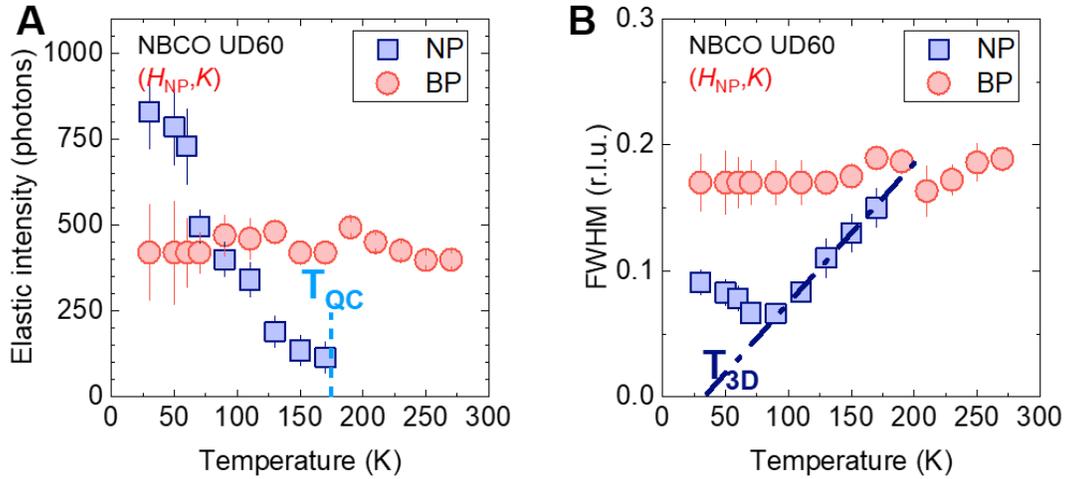

**Fig.S4**: **Broad and narrow peaks in sample UD60.** The plots summarize the parameters of the two Lorentzian profiles, used to describe the quasi-elastic peaks measured in sample UD60, along $K$ (squares refer to the narrow peak, circles to the broad peak). **(A)** Intensity vs temperature; **(B)** Width vs temperature. $T_{CDW}$ = 175 K. $T_{co}$ = 32.9 K, in perfect agreement with the $T_{co}$ determined from the FWHM(T) of the same sample, but along the $H$ scan (see both Fig. 3B and the blue square at $p$ = 0.11 in Figure 4A).

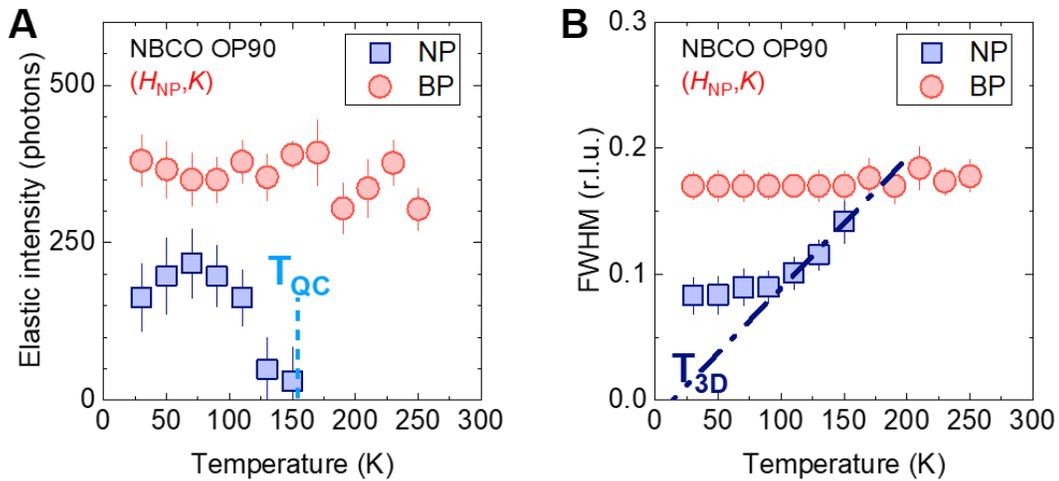

**Fig.S5**: **Broad and narrow peaks in sample OP90.** The plots summarize the parameters of the two Lorentzian profiles, used to describe the quasi-elastic peaks measured in sample OP90, along $K$ (squares refer to the narrow peak, circles to the broad peak). **(A)** Intensity vs temperature; **(B)** Width vs temperature. $T_{QC}$ = 155 K, while $T_{3D}$ = 14 K. The average of this $T_{3D}$ value with that determined from the FWHM(T) of the same sample, but along the $H$ scan (see Fig. 3D) corresponds to the blue square at $p$ = 0.17 in Figure 4A of the main manuscript. The intensity of the narrow peak along $K$ is lower than along $H$, and the width broader. This is due to the $K$ scan, performed slightly off from the maximum of the narrow peak.

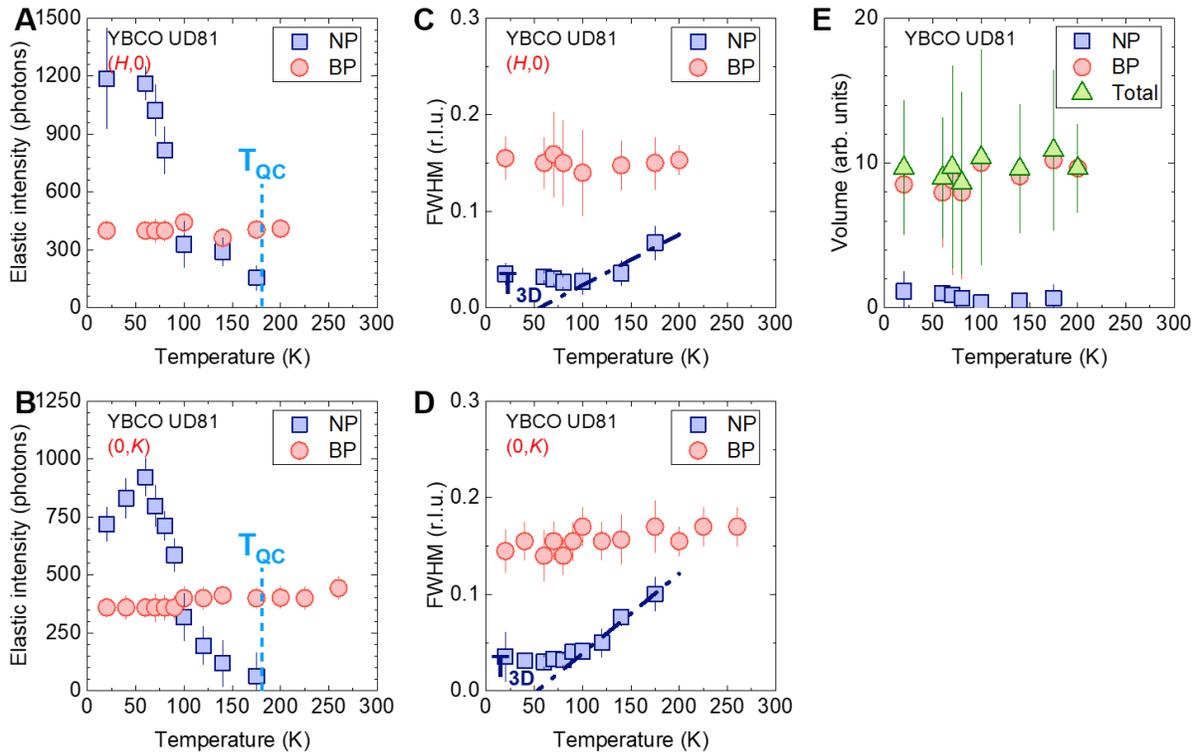

**Fig.S6: Broad and narrow peaks in sample UD81.** The plots summarize the parameters of the two Lorentzian profiles, used to describe the quasi-elastic peaks measured in sample UD81 (squares refer to the narrow peak, circles to the broad peak). Intensity vs temperature along $H$ **(A)** and along $K$ **(B)**. Width vs temperature along $H$ **(C)** and along $K$ **(D)**. $T_{CDW}$ = 180 K. The $T_{co}$ values are 55 K ($H$ scan) and 53 K ($K$ scan). The average of the two values corresponds to the blue square at $p$ = 0.14 in Figure 4A. **(E)** Volume as a function of the temperature for both the broad and the narrow peak.

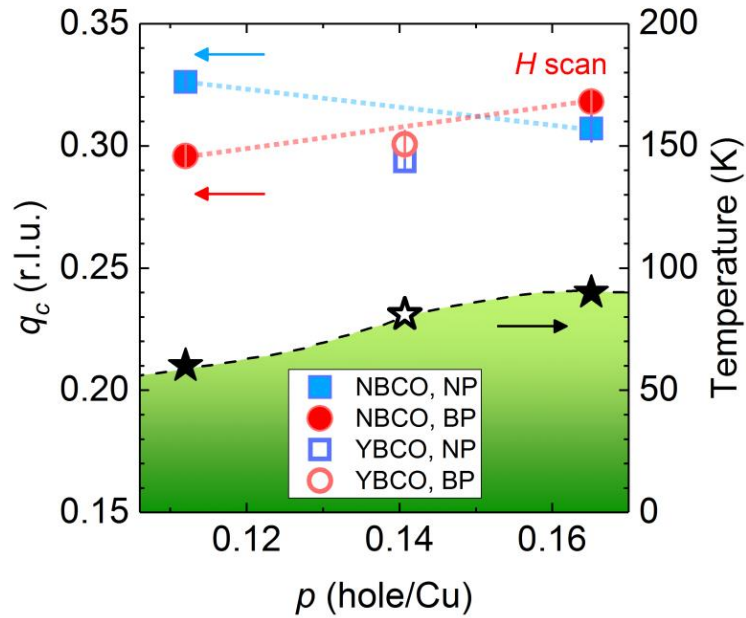

**Fig.S7: Doping dependence of $q_c$ for the broad and the narrow peak.** The figure confirms what already highlighted in Figures 2D-E of the main manuscript: the $q_c$ values of the BP and of the NP do not coincide (left axis). This has been observed for the three investigated samples. Even though different, the $q_c$ values of both the BP and NP are around 0.3 r.l.u. proper of the YBCO family, whose CO modulation is not linked to the spin order modulation. However, while the incommensurability of the narrow peak decreases with increasing $p$, as usually measured in these cases, the broad peak exhibits the opposite trend. The significantly high $q_c$ value measured for both the NP and the BP of sample OP90 might be due to the film being twinned, differently than the other two investigated samples. The three stars on the $T$-$p$ superconducting dome (right axis) refer to the critical temperatures of the investigated samples.

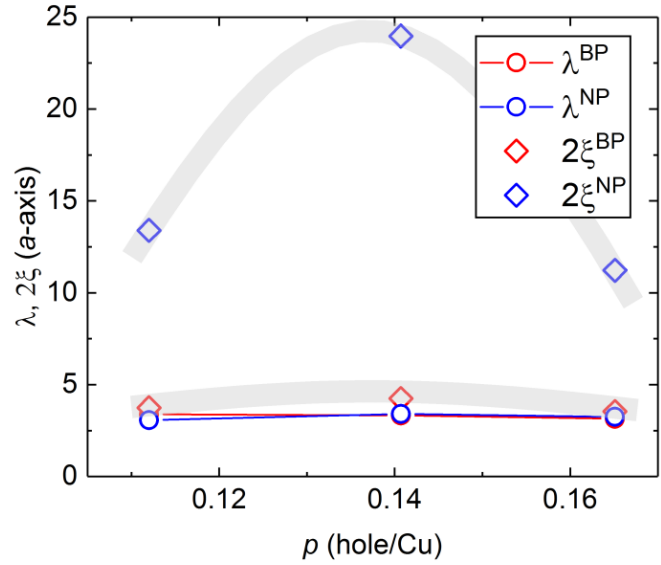

**Fig.S8: Correlation length ξ and wavelength λ of the charge order described by the broad (BP) and by the narrow (NP) peak**. The correlation length $\xi$ and the wavelength $\lambda$ have been determined, in units of the lattice parameter $a$, respectively as $(\pi \cdot w_{Tc})^{-1}$ and $(q^*)^{-1}$. Here, $w_{Tc}$ is the FWHM at $T = T_c$. In particular, $2\xi$ represents the diameter of the cluster in which charge order correlation is present. In the plot, $2\xi$ and $\lambda$ are shown, as a function of the doping $p$, for both NP and BP. The grey bands are guides to the eye.

As a consequence of the weak dependence of $q^*$ with $p$, the wavelength of the charge order is very similar at all doping for both the peaks, varying in the range $[3a, 3.4a]$. The correlation diameter $2\xi$ has also the same doping behaviour for both the peaks, being maximum for the sample at $p \approx 0.14$. However, while for the narrow peak $2\xi$ is much larger than $\lambda$, in the case of the broad peak $2\xi \approx 4a$. The size of the region, where the CDF (associated to the BP) is coherent, is comparable with a single period of the charge modulation. This result poses a severe limitation to a correlation reduction at temperatures higher than $T_c$, similarly to the case of the NP. A further increase of the width of the broad peak would indeed imply the loss of spatial coherence for the CDF.

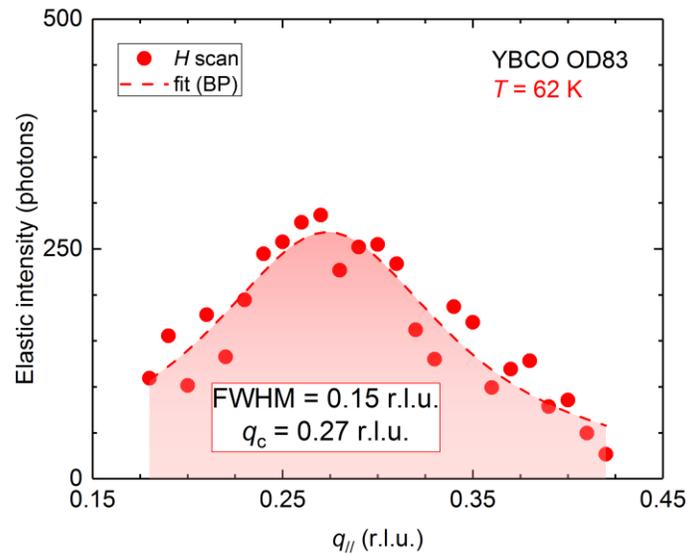

**Fig.S9: Scan of the quasi-elastic intensity along the *H* direction for a slightly overdoped YBCO film (OD83, $p \approx 0.18$).** Several RIXS spectra at different $q_{//}$ values have been measured at $T$ = 62 K, below the critical temperature of the film ($T_c \approx 83$ K). The quasi-elastic component of the spectra, integrated in the energy range [-0.2 eV, +0.15 eV] is plotted as a function of $q_{//}$ (red circles), after subtraction of the linear background. A clear peak is still present, which can be fitted by a Lorentzian profile (dashed line). The width of this peak is comparable to that of the broad peak, that we have associated to the CDF. At this doping value, since the NP - related to quasi-critical CDW - is absent, the broad peak is the only detectable signal already at low temperatures.

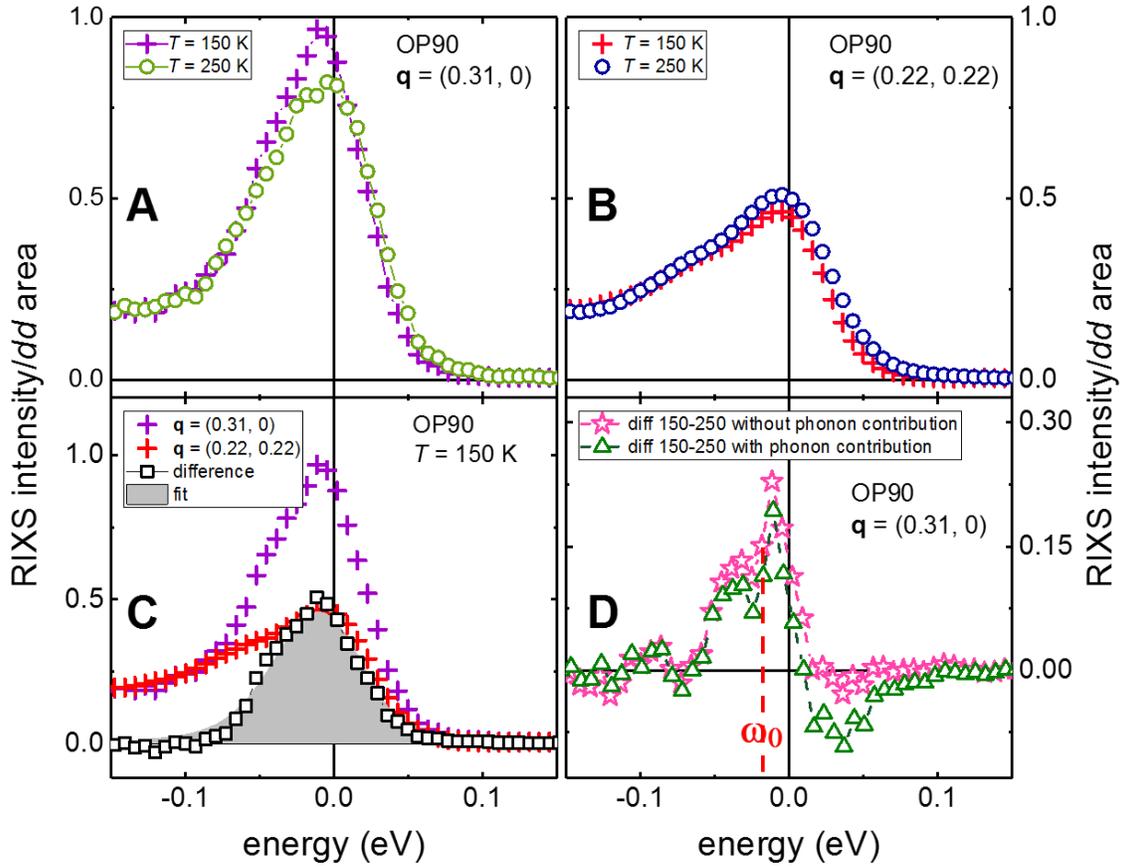

**Fig.S10: Subtraction of the phonon contribution. (A)** Quasi-elastic component of the spectra at $T$ = 150 and 250 K, measured on sample OP90 at $q_{//}$ = (0.31, 0). **(B)** Quasi-elastic component of the spectra at $T$ = 150 and 250 K, measured on sample OP90 at $q_{//}$ = (0.22, 0.22). Differently than along $H$, the intensity increases when increasing the temperature. This is mostly due to the increased population of the quasi-zero-energy-phonons at high temperature. **(C)** The quasi-elastic component of the spectra measured at $q_{//}$ = (0.31, 0) and $q_{//}$ = (0.22, 0.22) ($T$ = 150 K) are plotted together. Their difference (squares), which well approximates the CDW contribution within the spectrum, has been fitted considering the CDW instability theory (solid area). **(D)** The difference of the 150 K - 250 K spectra (stars), achieved after subtraction of the phonon contribution, is shown, together with the difference of the raw spectra, shown in panel (A). The two sets of data are very similar, and they can be both fitted considering a characteristic energy $\omega_0 \approx$ 15 meV.

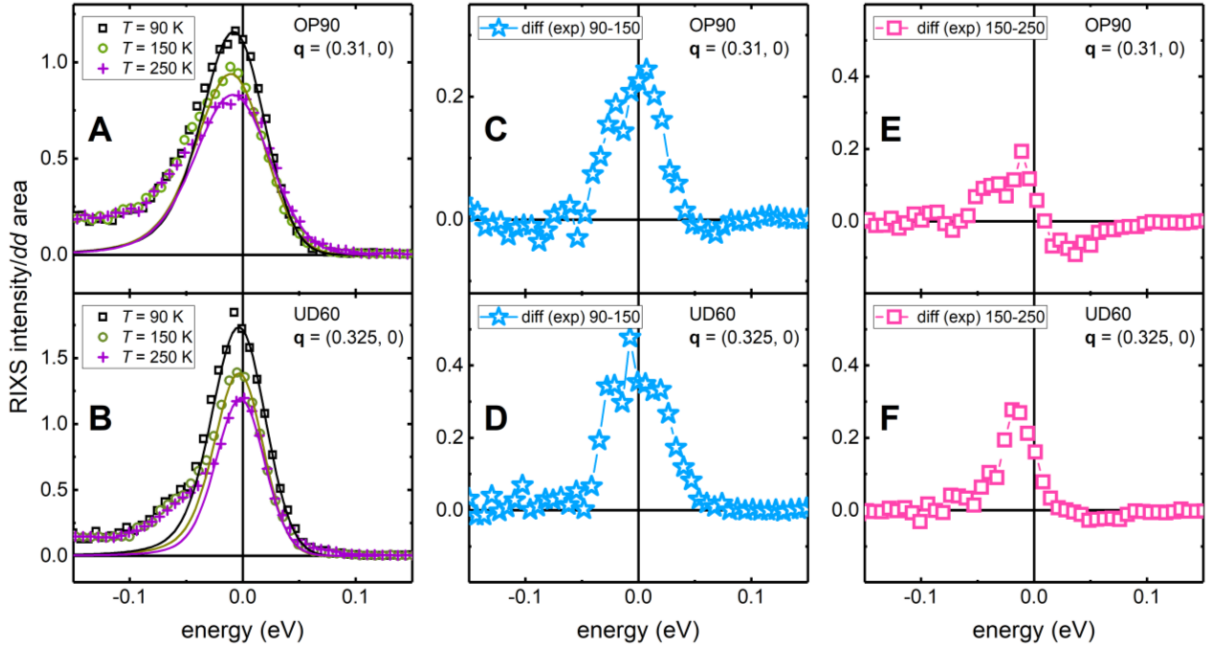

**Fig.S11: Doping and temperature dependence of the BP characteristic energy. (A)-(B)** Quasi-elastic component of the spectra, without subtraction of the phonon contribution, at $T$ = 90, 150 and 250 K, measured on sample OP90 and UD60. Together with the experimental data (symbols), also the curves used for the fit (lines) have been plotted. The data are in agreement with the theory assuming for sample OP90 $\omega_0 \approx 7$ meV at 90 K and $\omega_0 \approx 15$ meV at both 150 K and 250 K; for sample UD60 $\omega_0 \approx 3.3$ meV at 90 K and $\omega_0 \approx 5.5$ meV at both 150 K and 250 K. **(C)-(D)** The experimental differences of the 90 K - 150 K spectra are shown for both samples OP90 and UD60. **(E)-(F)** The experimental differences of the 150 K - 250 K spectra (squares) are shown for both samples OP90 and UD60.